\documentclass[emulateapj,natbib]{emulateapj}

\begin{document}
\title{Direct Imaging and Spectroscopy of a Candidate Companion 
Below/Near the Deuterium-Burning Limit 
In The Young Binary Star System, ROXs 42B}
\author{
Thayne Currie\altaffilmark{1}, 
Sebastian Daemgen\altaffilmark{1},
John Debes\altaffilmark{2},
David Lafreniere\altaffilmark{3},
Yoichi Itoh\altaffilmark{4},
Ray Jayawardhana\altaffilmark{1},
Thorsten Ratzka\altaffilmark{5},
Serge Correia\altaffilmark{6}
}
\altaffiltext{1}{Department of Astronomy and Astrophysics, University of Toronto, 50 St. George St., Toronto, ON, Canada
\email{currie@astro.utoronto.ca}}
\altaffiltext{2}{Space Telescope Science Institute, Baltimore, MD}
\altaffiltext{3}{D\`epartment de Physique, Universit\'e de Montreal, Montr\'eal, QC, Canada}
\altaffiltext{4}{Nishi-Harima Observatory, University of Hyogo, Kobe, Japan}
\altaffiltext{5}{Universit\"ats-Sternwarte M\"unchen, Ludwig-Maximilians-Universit\"at, M\"unchen, Germany}
\altaffiltext{6}{Institute for Astronomy, University of Hawaii, Pukalani, HI}
\begin{abstract}
We present near-infrared high-contrast imaging photometry and integral field spectroscopy of ROXs 42B, a binary 
M0 member of the 1--3 Myr-old $\rho$ Ophiuchus star-forming region, from data collected over 7 years.
  Each data set reveals a faint
companion -- ROXs 42Bb -- located $\sim$ 1.16\arcsec{} ($r_{proj}$ $\approx$ 150 $AU$) from the primaries at a 
position angle consistent with a point source identified earlier by \citet{Ratzka2005}.
ROXs 42Bb's astrometry is inconsistent with a background star but consistent with a bound companion, possibly 
one with detected orbital motion.
The most recent data set reveals a second candidate companion at $\sim$ 0\farcs{}5 of roughly equal brightness, 
though preliminary analysis indicates it is a background object.  
ROXs 42Bb's $H$ and $K_{s}$ band photometry is similar to dusty/cloudy young, low-mass late M/early L dwarfs.
$K$-band VLT/SINFONI spectroscopy shows ROXs 42Bb to be a cool substellar object 
(M8--L0; $T_{eff}$ $\approx$ 1800--2600 $K$), not a background dwarf star, with a spectral shape indicative of
young, low surface gravity planet-mass companions.
We estimate ROXs 42Bb's mass to be 6--15 $M_{J}$, either below the deuterium burning limit and thus 
planet mass or straddling the deuterium-burning limit nominally separating planet-mass companions 
from other substellar objects.  Given ROXs 42b's projected separation and 
mass with respect to the primaries, it may represent the lowest mass objects formed like binary stars or a class of 
planet-mass objects formed by protostellar disk fragmentation/disk instability, the latter slightly blurring the 
distinction between non-deuterium burning planets like HR 8799 bcde and low-mass, 
deuterium-burning brown dwarfs.
\end{abstract}
\keywords{planetary systems, stars: individual: ROXs 42B} 
\section{Introduction}
In the last decade, high-contrast \textit{direct imaging} observations of young stars 
have provided a way to identify, confirm, and study the properties of self-luminous planet-mass 
companions around nearby stars undetectable 
by radial-velocity and transits \citep[e.g.][]{Chauvin2004,Marois2008,Marois2011,
Lagrange2010,Lafreniere2008,Lafreniere2010,
Currie2012a,Kalas2008,Ireland2011,Carson2013,Rameau2013}.  These companions are typically found at wide ($r$ $\sim$ 25-300 $AU$) separations.
Besides Fomalhaut b \citep[M $<$ 2 $M_{J}$][]{Currie2012a}, they
 have super jovian ($M$ $\approx$ 4--15 $M_{J}$) masses.  Thus, these objects 
significantly challenge even advanced models for 
forming them as planets (by core accretion or disk instability) or as very low-mass binary 
companions \citep[e.g.][]{KenyonBromley2009,Kratter2010,Boss2011}.  

 Thus far, these companions can be divided into two groups: 1) low companion-to-star mass ratio objects like 
HR 8799 bcde and $\beta$ Pic b comprising the extrema of a population 
extending to planets detected by radial-velocity and plausibly formed by core accretion 
and 2) high mass ratio companions (e.g. GSC 06214B) comprising the tail of a population
extending up to brown dwarf masses ($M$ $\gtrsim$ 13-15 $M_{J}$) and formed by other means \citep{Currie2011a,Janson2012a}.
Some companions have a more ambiguous nature due to their 
uncertain ages (and, thus, uncertain masses/mass ratios) \citep[e.g.][]{Carson2013}.
Companions with well-defined ages could better probe the extremes of these populations,
clarifying or blurring the distinction between bona fide planets and the lowest mass 
brown dwarfs.

In this Letter, we report direct imaging and spectroscopy of a candidate 
6--15 $M_{J}$ companion to ROXs 42B, a binary M star and likely member of 
the $\rho$ Ophiuchus star-forming region, that may add to the 
list of rare wide separation, young imaged planet-mass companions.  

\section{Basic System Properties and Data}
 ROXs 42B was originally identified as
 an M0, possibly x-ray emitting star \citep{Bouvier1992}\footnote{M0V spectroscopic standards 
used at the time of ROXs 42B's discovery draw from many sources
(\url{http://www.pas.rochester.edu/$\sim$emamajek/spt/M0V.txt}).   
Uncertainties in which calibrator was used for ROXs 42B imply
  a conservative spectral type range of 
K5--M1 (E. Mamajek 2013, pvt. comm.).} and member of the 1--3 $Myr$
 $\rho$ Ophiuchus star-forming region \citep[][]{Luhman1999,Wilking2008,Erickson2011} 
located between the L1888 and L1989N/S clouds. 
Lunar occulation measurements showed it to be a close
binary ($r$ $\sim$ 0\farcs{}05) \citep{Simon1995} where the brighter (M0) component 
contributes two-thirds of the system brightness \citep{Ratzka2005}.
Sensitive optical to (sub)millimeter data reveal no clear
 evidence for circumstellar gas or dust \citep[e.g.][]{Cieza2007,Andrews2007}.

We initially identified a candidate 3rd member of the ROXs 42B system located at a projected separation of 
$\approx$ 1\farcs{}16 as a part of a general search for planet/brown 
dwarf companions in archival Keck, Subaru, and VLT data \citep[][]{Currie2012b,Currie2013a}.
As we later found, \citet{Ratzka2005} identified the same 3rd member as a point source 
from speckle imaging data obtained in 2001.  
Although they did not estimate the object's mass, their $K$-band flux ratio (0.002 $\pm$ 0.001) implied a mass
below the deuterium-burning limit given the age and distance of $\rho$ Oph and standard hot-star planet evolution 
models \citep[e.g.][]{Baraffe2003,Spiegel2012}, spurring us to rereduce additional 
archival data to confirm or reject its status as a bound companion and determine from multiwavelength 
photometry and spectroscopy whether its SED is consistent with 
those for other young directly-imaged planetary-mass companions 
\citep[e.g.][]{Lafreniere2008,Lafreniere2010,Currie2011a,Barman2011}.

Our combined data set (Table 1) includes Keck/NIRC2 and Subaru/CIAO near IR
imaging and VLT/SINFONI integral field spectroscopy (IFS) and astrometry from \citet{Ratzka2005}.
The Keck/NIRC2 and Subaru/CIAO data we consider were taken in three epochs between 2005 and 2011 
in the $H$, $K$ or $K_{s}$ broadband filters.  The first Keck epoch data 
and CIAO data were obtained with (extremely) modest AO corrections,
 while the 2011 Keck and VLT images were nearly diffraction limited. 
  All data were obtained in classical imaging, not 
\textit{angular differential imaging} \citep{Marois2006}, and in various dither/nod patterns to remove 
the sky background.  The field of view was $\approx$ 10, 20, and 3 arc-seconds on a side
 for the NIRC2, CIAO, and SINFONI data, respectively.  The SINFONI spectral resolution 
was R $\approx$ 1500.  The primary in each data set was unsaturated.

Basic image processing steps for the photometric data followed those taken for 
IR data that were likewise obtained in a dither pattern \citep[e.g.][]{Currie2011a}.  After sky subtraction, 
identifying and removing hot/cold/bad pixels, correcting each image for distortion (for NIRC2 data 
only) and copying each to larger blank image, we registered the images to a common center following 
standard methods used before for centroiding unsaturated PSF cores.
Finally, we subtracted off a 2D radial profile of the primary to remove the halo light from each 
image, median-combined them, and rotated the combined image to the north-up position\footnote{The 
companion is well outside most of the halo light in each photometric data set and just slightly 
contaminated in the IFS data.   Compared to, say, HR 8799 bcde and $\kappa$ And b, 
it is at modest contrasts.
Thus, we did not need to perform sophisticated image processing techniques \citep[e.g.][]{Lafreniere2007a,Currie2012a}}. 

Using the EsoRex\footnote{http://www.eso.org/sci/software/cpl/esorex.html} 
pipeline, we reduced the SINFONI IFS data, performing sky subtraction, combining
the data into a spectroscopic cube, correcting for 
differential atmospheric refraction and curvature in the spectra.
We extracted the spectra using a 3-pixel radius annulus using 
QFitsView\footnote{www.mpe.mpg.de/\~ott/QFitsView/}. We applied the same reduction and 
extraction routine to a B9 telluric standard star 
observed during the same night at similar airmass and with the same instrumental 
setup. Its extracted spectrum was divided by a blackbody curve according to its effective 
temperature and its Brackett-$\gamma$ absorption feature at $\lambda\approx2.16\,\mu$m replaced 
by a straight line along the continuum \footnote{We mitigated contamination of the companion's spectrum 
by the primary's diffraction spike and halo by
defining sky regions along the diffraction spike on both sides of the companion.}.
Finally, we divided the extracted spectra of ROXS 42B and ROXs 42Bb by the corrected 
telluric standard.

\begin{figure}
\centering
\includegraphics[scale=0.315,trim=2mm 2mm 5mm 1mm,clip]{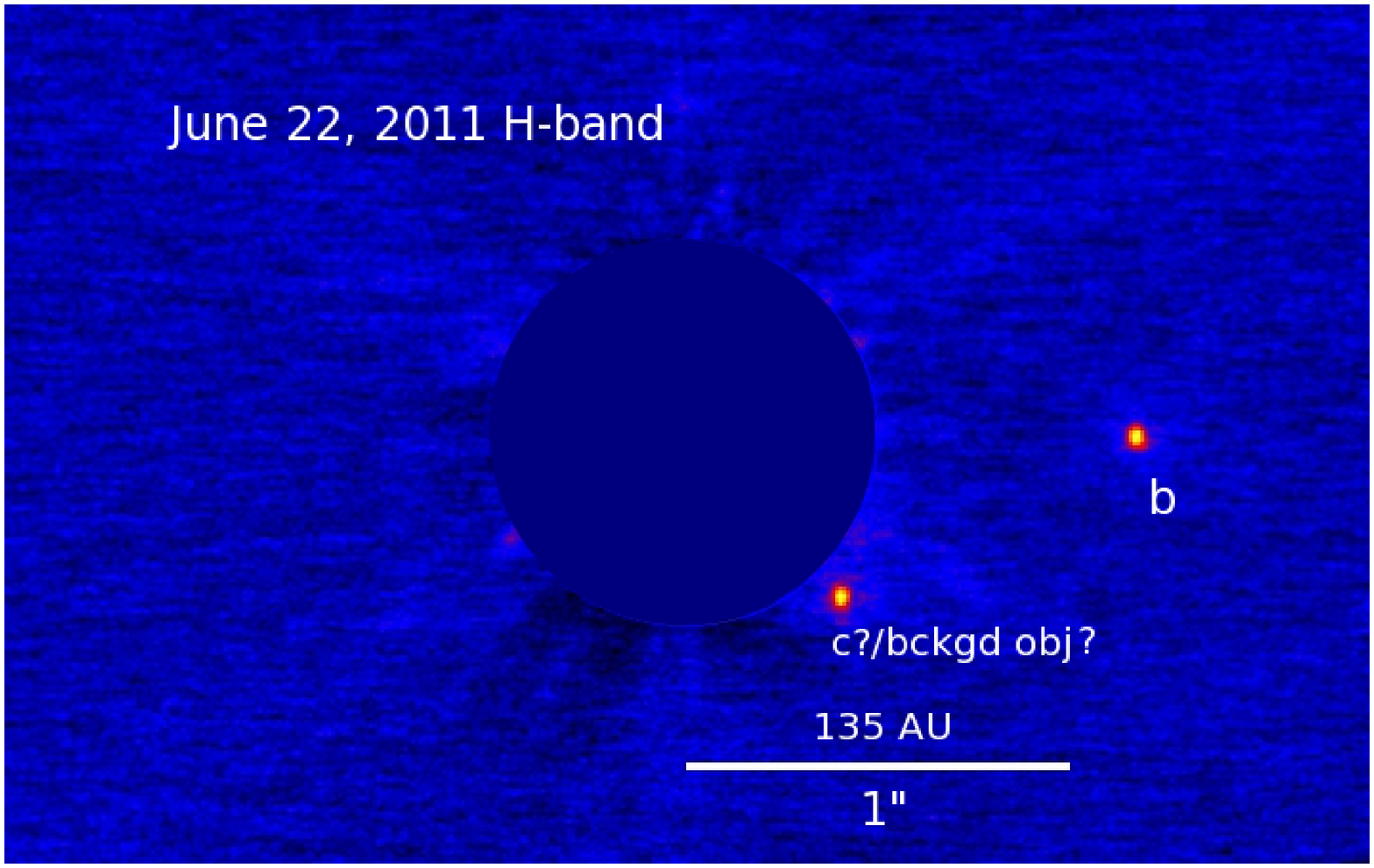}
\includegraphics[scale=0.34,trim=2mm 8mm 15mm -10mm,clip]{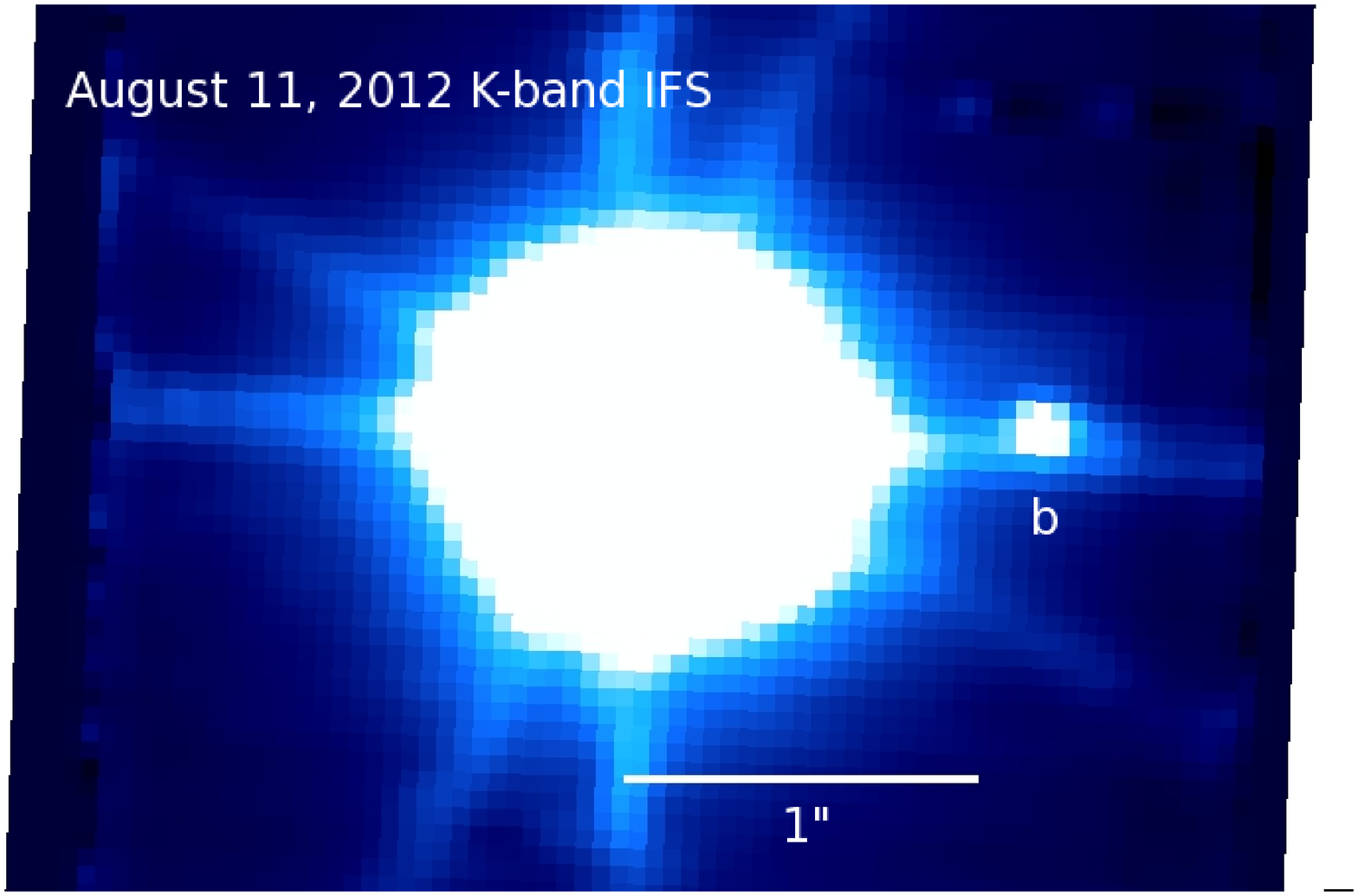}
\\
\includegraphics[scale=0.33,trim=1mm 1mm 1mm 1mm,clip]{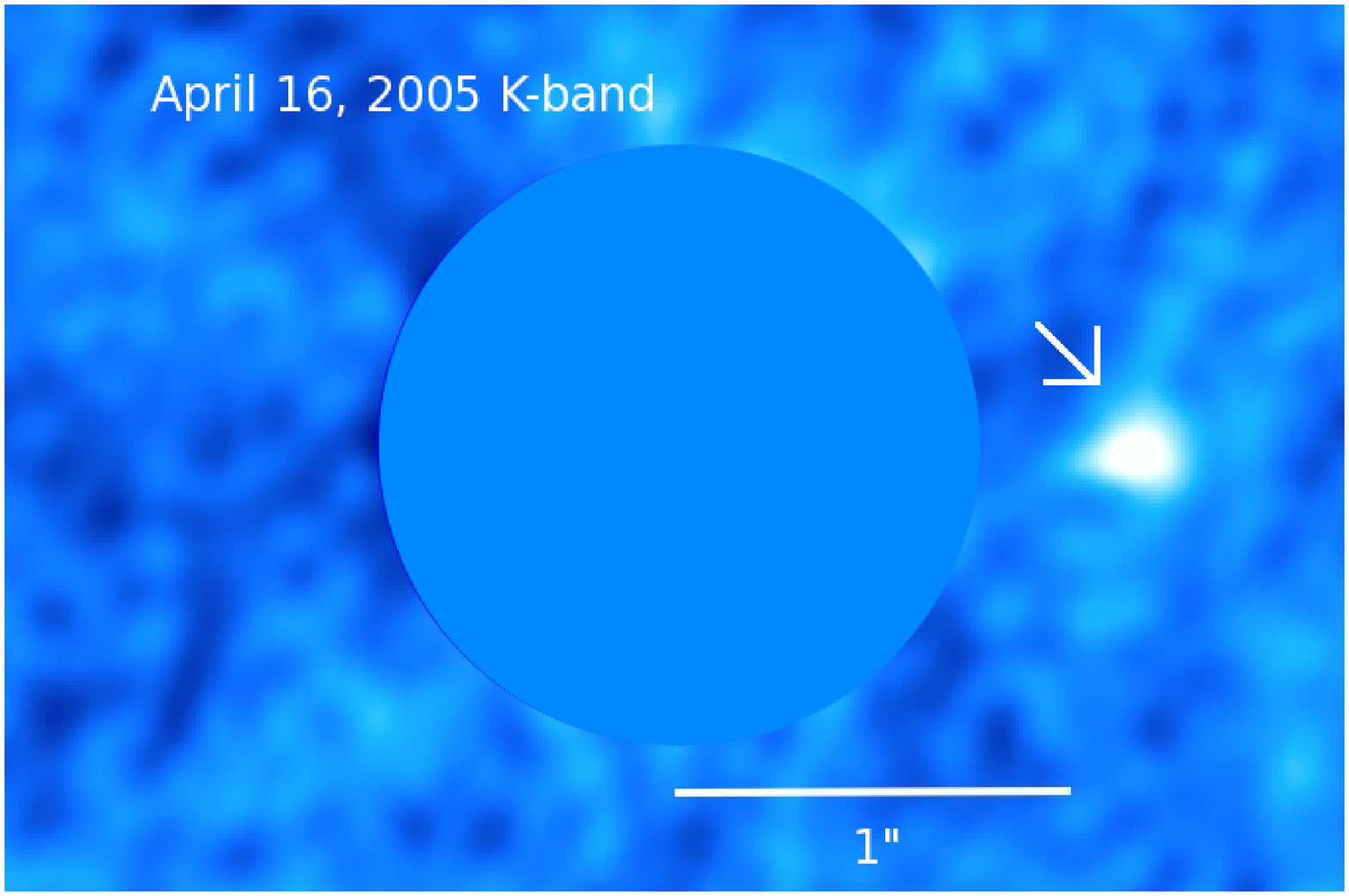}
\includegraphics[scale=0.33,clip]{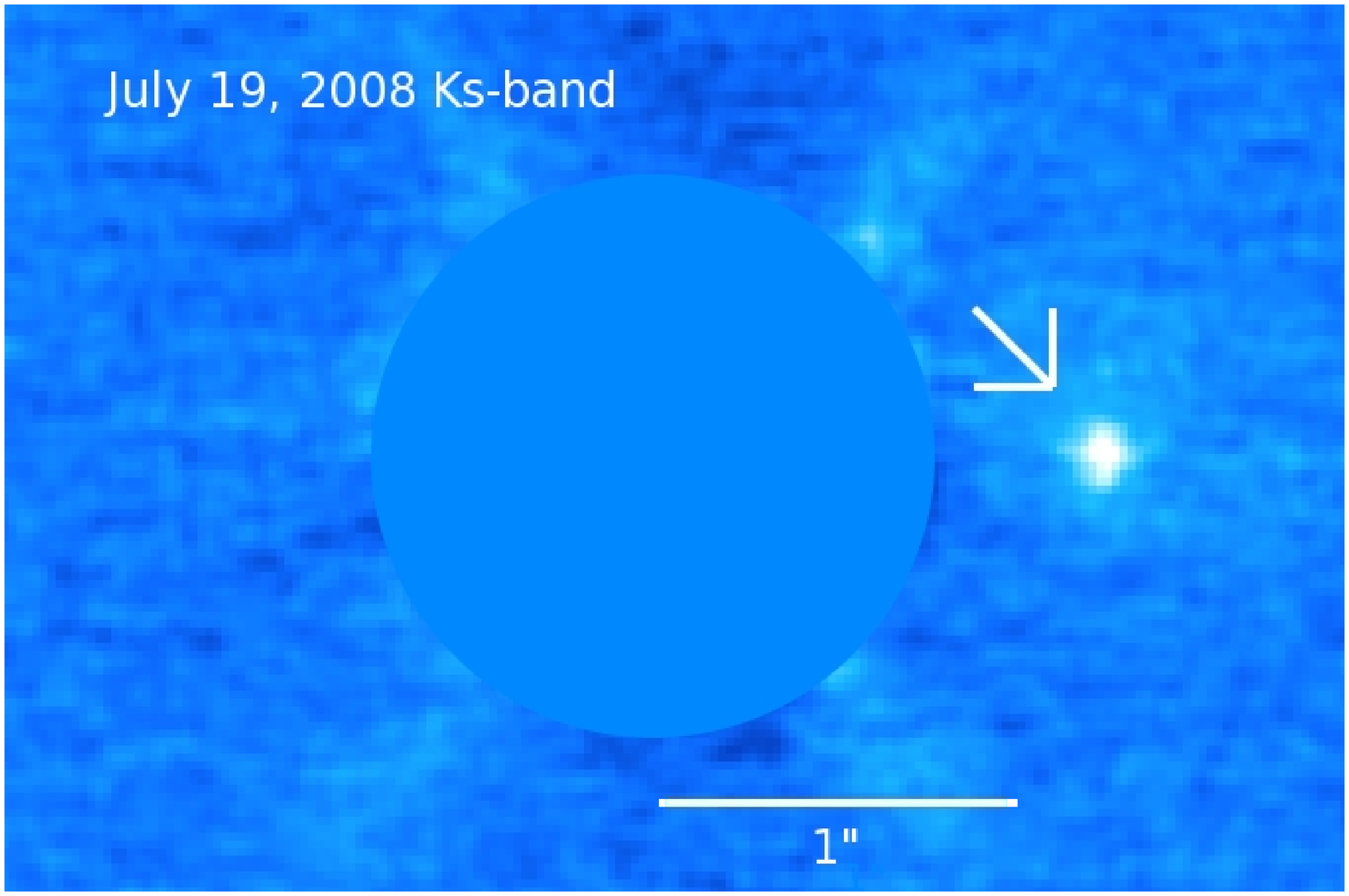}
\caption{Images of ROXs 42B from 2011 $H$-band NIRC2 data (top-left), 2012 VLT/SINFONI 
$K_{s}$ band IFS data (top-right, collapsed cube shown), 2005 $K$-band Keck/NIRC2 data 
(bottom-left), and 2008 Subaru/CIAO data (bottom-right).  The 2011 NIRC2 and 2012 SINFONI 
data are not smoothed; the older data sets are convolved with a gaussian kernal equal to the 
image full-width half-maximum of the primary star.  All images are rotated 'north-up'.}
\label{images}
\end{figure}

Figure \ref{images} displays each final image from our photometric data and the collapsed cube from the 
VLT/SINFONI spectroscopy, revealing a candidate companion, ROXs 42Bb, at $r$ = 1\farcs{}15--1\farcs{}17 
(labeled or denoted with an arrow, 3 o'clock position in each panel) at a signal-to-noise ratio greater than 5.
Assuming a distance of 135 $pc$ \citep{Mamajek2008}, this angular separation corresponds to a 
projected physical separation of $\approx$ 150 $AU$.
Furthermore, the $H$-band 2011 data set reveal a second candidate 
at $r$ $\sim$ 0\farcs{}5 ($\approx$ 65 $AU$), which has roughly the same brightness 
as ROXs 42Bb.  The data quality in the other epochs is not sufficient to recover 
this object.  We do not resolve the two primary stars in any data set.

\section{Analysis}
\subsection{Astrometry}
We use gaussian fitting (the $IDL$ $Astrolib$ subroutine \textrm{gcntrd.pro}) to determine the position of 
ROXs 42Bb in each photometric data set, assuming a FWHM equal to that estimated for the star,
 and the \textit{DAOFIND} centroiding routine for the SINFONI IFS data.
For NIRC2 astrometric calibration, we adopt the pixel scale and north position angle offset 
(9.952 mas pixel$^{-1}$, $PA$ = -0.252$^{o}$) for narrow camera data from \citet{Yelda2010}.
For CIAO data, we adopt a pixel scale of 21.33 mas pixel$^{-1}$ \citep{Itoh2005} and a 
north position angle offset of 1.14$^{o}$.

\begin{figure}
\centering
\includegraphics[scale=0.3]{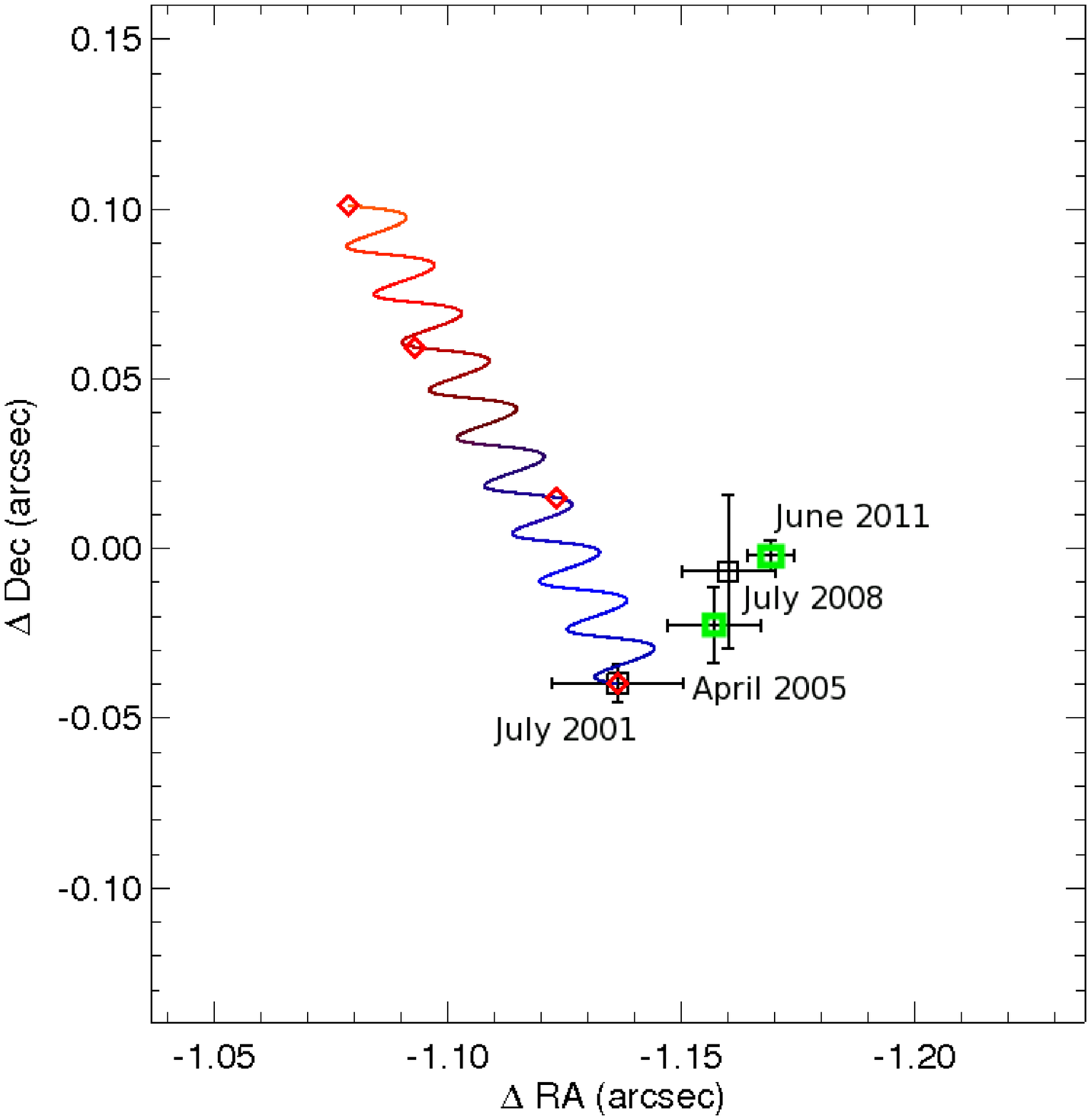}
\includegraphics[scale=0.4]{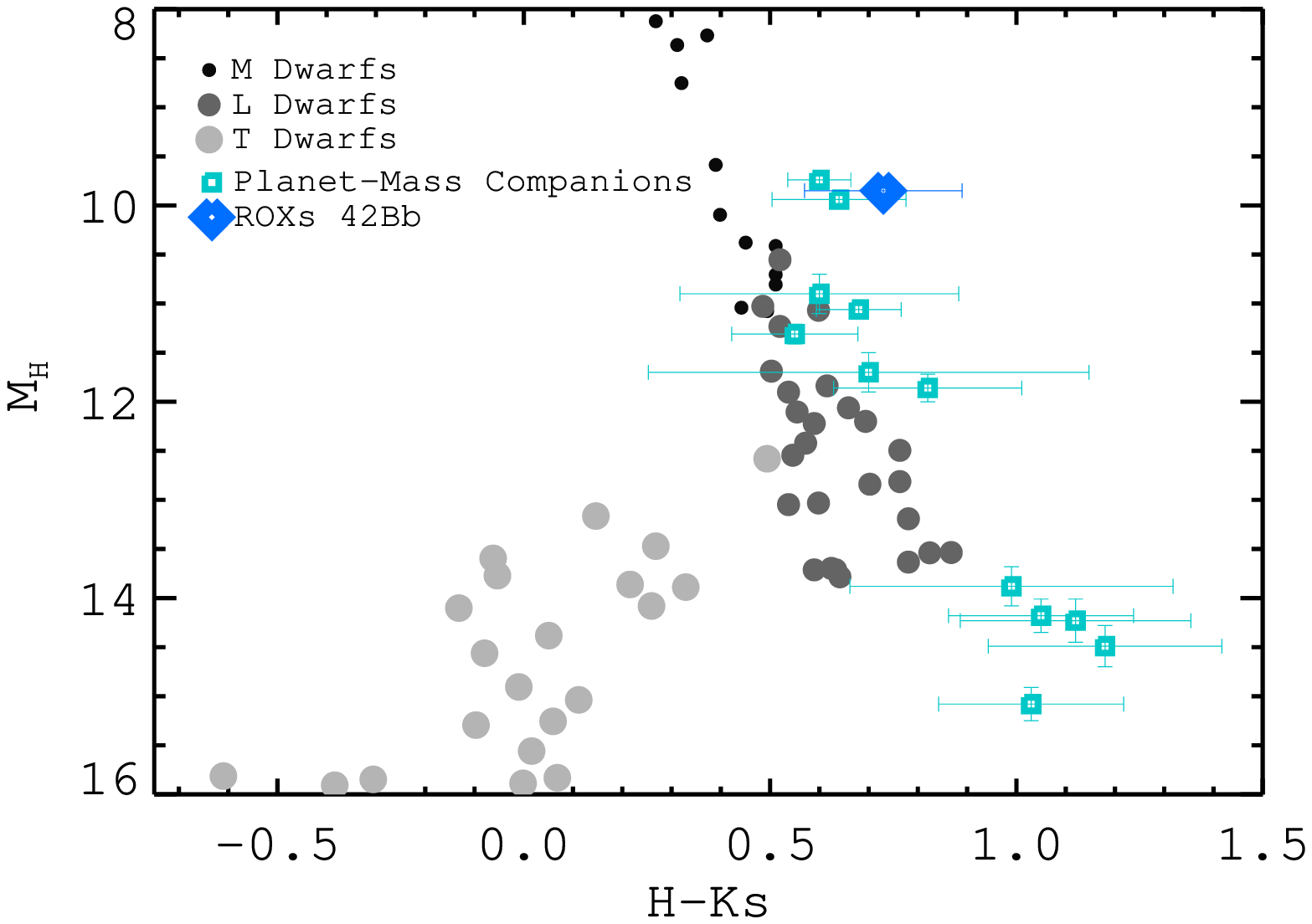}
\caption{
(Left) ROXs 42Bb's 
astrometry (black squares) from 2001 to 2011 and between just the two Keck/NIRC2 epochs (overplotted green squares) 
is inconsistent with a background star's motion (red squares).  (Right) ROXs 42Bb's $H$/$H$-$K_{s}$ 
position is consistent with cloudy/dusty, L type low surface gravity/mass substellar objects}.
\label{roxs42b_initan}
\end{figure}

Figure \ref{roxs42b_initan} (left panel) compares ROXs 42Bb's astrometry derived from 
CIAO and NIRC2 data to the motion 
expected for a background object assuming a parallax of 7.4 mas.  We adopt the 
UCAC4 \citep{Zacharias2012} catalog's proper motion for ROXs 42B: $\mu_{\alpha}$ $cos(\delta)$, $\mu_{\delta}$ = 
-5.9 $\pm$ 1.6 mas yr$^{-1}$, -14.1 $\pm$ 1.8 mas yr$^{-1}$.  The motion vector
predicted for a background star is inconsistent with ROXs 42Bb's measured astrometry.  The 
predicted change in position between 2005 and 2011 is $\sim$ $\Delta(\alpha, \delta)$ $\approx$ 
[40 mas, 100mas] while our astrometry is nearly consistent with no motion at all:
the Keck data alone reject a background star hypothesis at more than the 5-$\sigma$ level.  
The 2012 SINFONI data are undersampled such that deriving astrometry with a precision on par 
with NIRC2 and CIAO data is challenging.  Nevertheless, our derived position from these 
data using the fits header information ($r$ $\sim$ 1\farcs{}15 $\pm$ 0\farcs{}03, $PA$ $\sim$ 
270$^{o}$ $\pm$ 2$^{o}$) is consistent with the 2011 Keck astrometry.  

A free-floating substellar object at the location of ROXs 42Bb with a similar 
space motion would be a rare occurrence.
  Even assuming twice as many young stellar objects as listed in \citet{Wilking2008} 
in $\approx$ 29 square degree area covering $\rho$ Oph 
and 10 times as many substellar objects as stellar members
instead of an equal amount \citep{Marsh2010} or 
fives times fewer \citep[][]{Muzic2012} and that all objects have proper motions 
indistinguishable from ROXs 42B, the probability of 
contamination on the 10\arcsec{}x10\arcsec{} Keck/NIRC2 field is 
$\approx$ 5$\times$10$^{-3}$.  

Although astrometry obtained with the same instrument setup (the 2005 and 2011 NIRC2 data) 
is formally consistent with no motion considering errors, the astrometric drift from epoch to epoch 
is also consistent with orbital motion.  Between the two Keck/NIRC2 epochs ($\Delta$t $\sim$ 
6 years), ROXs 42Bb changes position by $\approx$ 0\farcs{}0197, yielding an apparent space motion of 
$\approx$ 3.2 mas/year.  For a planet at 150 $AU$ on a circular orbit with a primary mass of 
$\approx$ 1 $M_{\odot}$, the mean positional change can be up to 3.8 mas/year for a face-on orbit.
While the relative positional changes for ROXs 42Bb suggests its orbit may not be viewed face-on, 
the object could be near periastron, admitting a larger apparent motion.

\subsection{Photometry and Spectroscopy}
 ROXs 42B is unsaturated in each image, has precise \textit{2MASS} photometry ($J$ = 9.906 $\pm$ 0.02, 
$H$ = 9.017 $\pm$ 0.02, $K_{s}$ = 8.671 $\pm$ 0.02), and is not known to be variable.
Therefore, we flux-calibrated ROXs 42Bb from the primary fluxes and our 
contrast measurements (Table 1).

We adopt the
Keck photometry to yield $m_{H}$ = 15.88 $\pm$ 0.05 and $m_{K_{s}}$ = 15.00 $\pm$ 0.15.

Following \citet{Currie2011a,Currie2013b}, we compare
 the near IR photometry of ROXs 42Bb to the sample of L/T dwarfs compiled by
\citet{Leggett2010} and directly-imaged planet-mass companions (defined as $M$ $\le$ 15 $M_{J}$; see 
values listed in Currie et al. 2013b).
Prior to these analyses, we derive the extinction of 
$A_{V}$ $\sim$ 1.9 from comparing the primary's $V$ and $K$-band photometry \citep[c.f.][]{Cieza2007} 
to predicted values for pre-main sequence M0 stars \citep{Pecaut2013}, adopting the 
extinction law from \citet{Cardelli1989} with $R_{V}$ = 3.1.  Then we deredden the companion and 
 derive absolute $H$ and $K$-band magnitudes: $M_{H}$ = 9.87 $\pm$ 0.05, 
$M_{Ks}$ = 9.13 $\pm$ 0.15.

The right panel of Figure \ref{roxs42b_initan} compares ROXs 42Bb to these populations.  
The companion's $H$-band luminosity is most consistent with late M to early L dwarfs.
  However, like planet-mass companions displayed here, it has a redder $H$-$K_{s}$ color 
than the field sequence, a feature explainable by dustier atmospheres usually associated 
with low mass/surface gravity objects \citep[c.f.][]{Burrows2006,Currie2013a}.
ROXs 42Bb shares a very similar color-magnitude diagram position to GSC 06214B and 
USco CTIO 108B \citep{Ireland2011,Bejar2008}, 
two substellar objects with masses of $\approx$ 6--16 $M_{J}$ and 10--15 $M_{J}$, respectively 
located in the nearby, slightly older Upper Scorpius association.
If ROXs 42B is a member of $\rho$ Oph or at least is similarly aged, it is likely lower mass than GCS 06214B and USco CTIO 108B.

\begin{figure}
\centering
\includegraphics[scale=0.45]{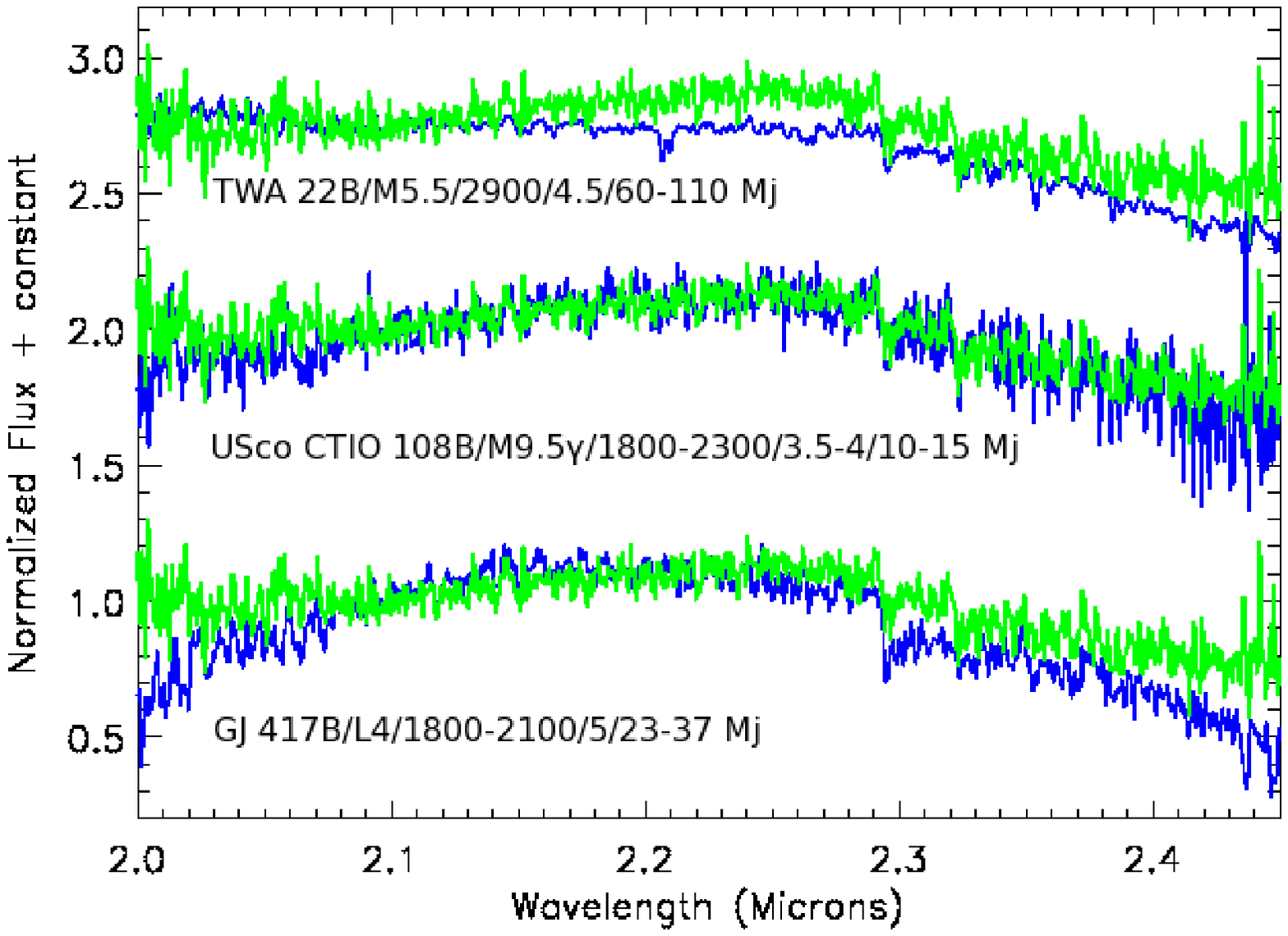}
\includegraphics[scale=0.45]{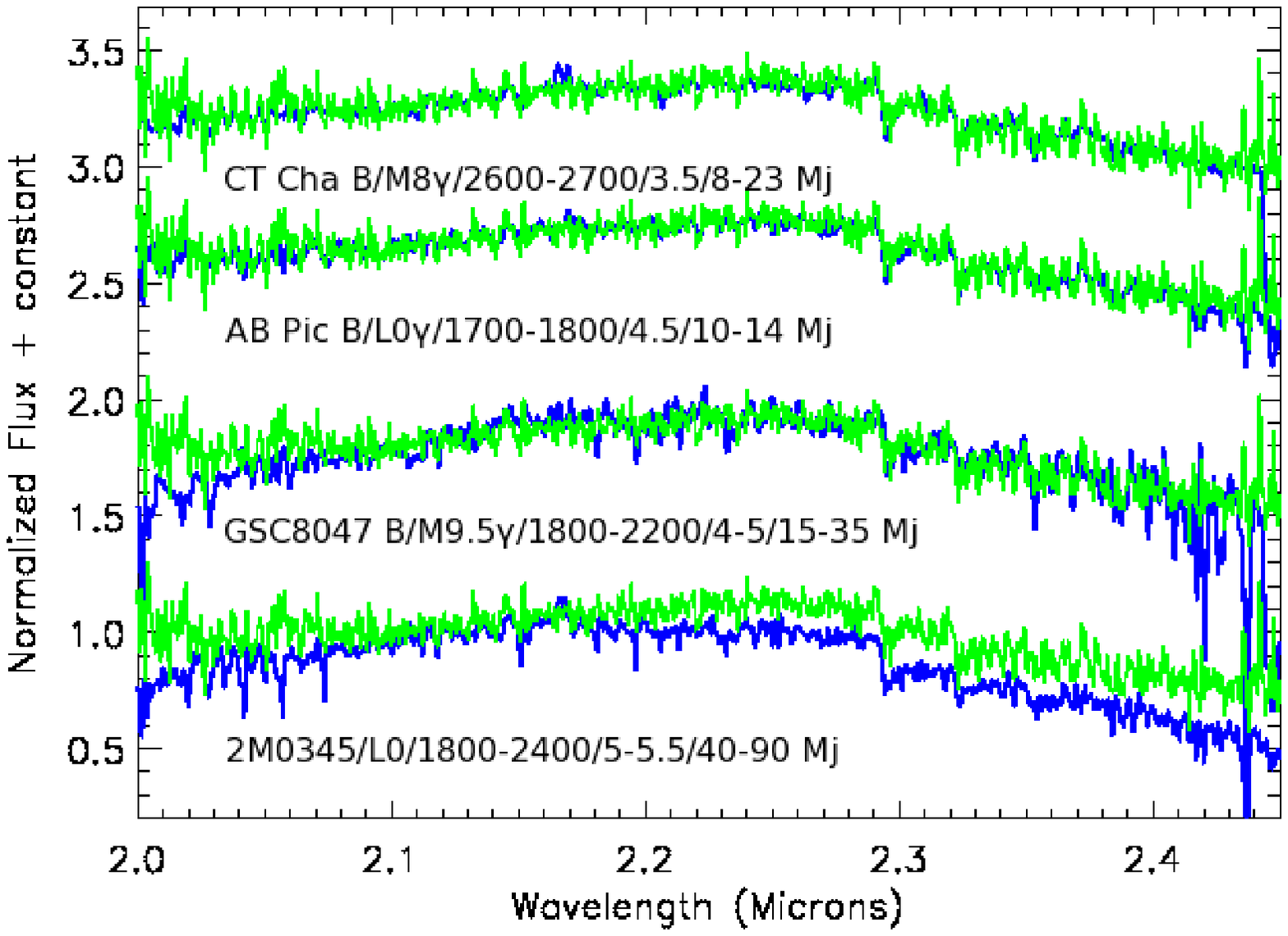}
\\
\includegraphics[scale=0.45]{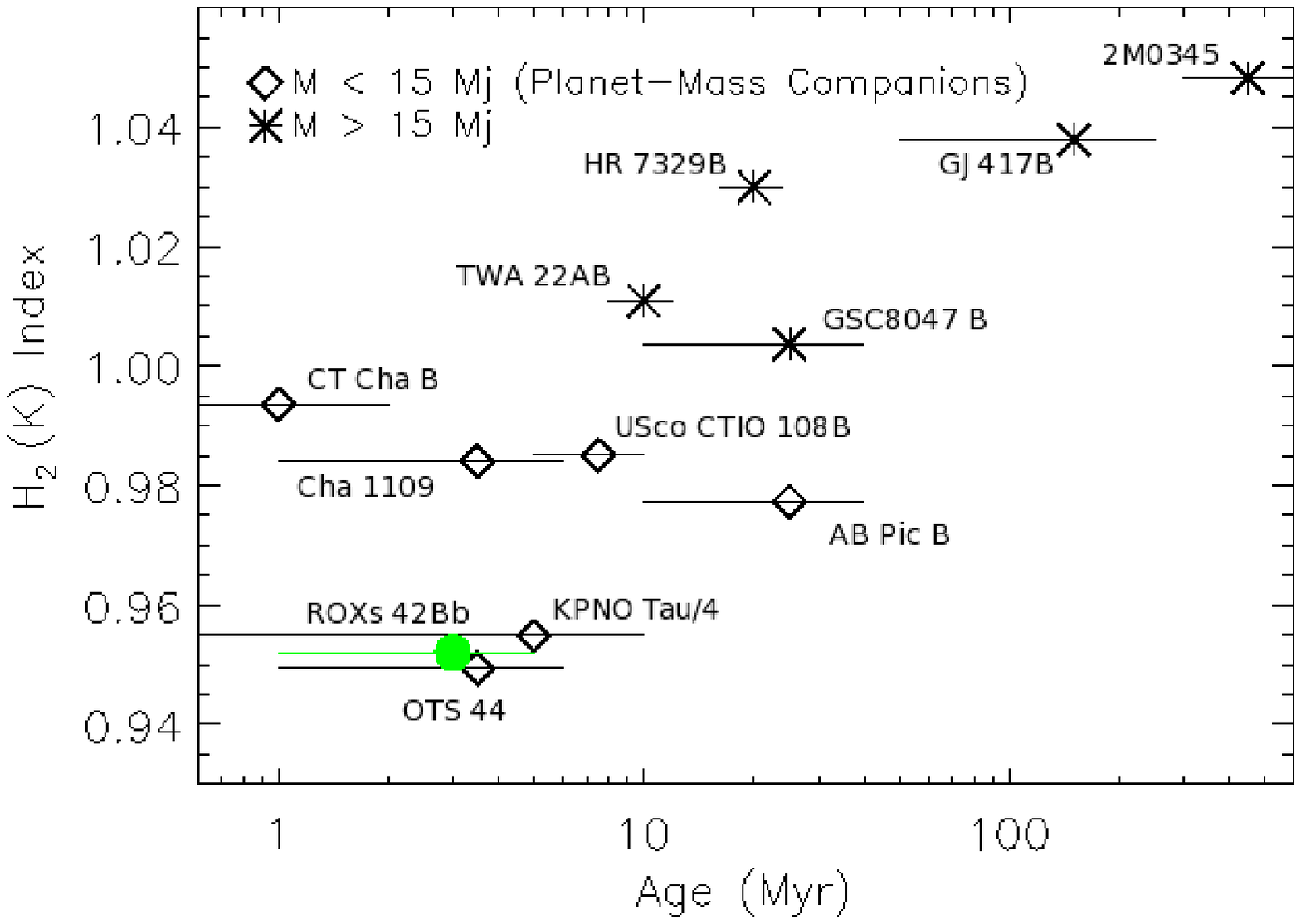}
\caption{
ROXs 42Bb spectrum (green) compared to other mid M to mid L dwarfs (blue): the target labels 
list the spectral type, $T_{eff}$, log(g), and mass (in units of jovian masses).  (left) The shape of the continuum is 
best matched by an objects near the M/L dwarf boundary that (right) are young, have low surface gravities, and masses 
below/at the deuterium-burning limit (e.g. AB Pic B, CT Cha B).  
(bottom) ROXs 42Bb's $H_{2}$(K) index compared to other objects in the \citet{Bonnefoy2013b} library.
}
\label{roxs42b_spec}
\end{figure}

Our SINFONI $K$-band spectrum of ROXs 42Bb (Figure \ref{roxs42b_spec}, top panels), when compared to spectra 
for M and L dwarfs from the \citet{Bonnefoy2013b} SINFONI library, provides even stronger evidence 
that the companion is most similar to young objects 
below/near the deuterium-burning limit.  Its spectral shape precludes it from being a stellar 
companion earlier than $\approx$ M5--M6 or a foreground mid L dwarf but appears consistent with 
at least some objects near the M/L dwarf transition.  

Furthermore, ROXs 42Bb's spectrum strongly resembles that for substellar, 
possibly planet-mass companions CT Cha B and AB Pic B, which have spectral 
types M8$_{\gamma}$ and L0$_{\gamma}$, respectively (Figure \ref{roxs42b_spec}, top-right panel).  Compared to GSC0847 B, 
which has a similar spectral type (M9.5$_{\gamma}$) but slightly higher inferred mass (15--35 $M_{J}$), 
the ROXs 42Bb deviates at $\lambda$ $\sim$ 2--2.1 $\mu m$.
The spectrum deviates from the field, higher surface gravity (log(g) = 5--5.5) L0 dwarf 
2M0345.  

The flatness of the $K$-band spectrum (the $H_{2}$(K) index) 
at 2.17--2.24 $\mu m$ may be a gravity/age diagnostic \citep[][e.g. ]{Luhman2004}\footnote{See also 
\citet{Allers2007} for gravity-sensitive features at $J$ and $H$ band}.
ROXs 42Bb's $H_{2}$(K) index ($\sim$ 0.95), as defined in \citet{Canty2013}, is smaller than all 
but one \citeauthor{Bonnefoy2013b} library objects, clearly distinguishable from that for 
objects more massive than $M$ $\sim$ 15 $M_{J}$ (bottom panel).  Thus, ROXs 42Bb's $K$-band 
spectral shape provides additional evidence that the companion is young and has a low mass/surface gravity.
From properties of the best-fitting comparison spectra, we estimate 
the following most-likely atmospheric properties for ROXs 42Bb: a spectral type of M8$_{\gamma}$--L0$_{\gamma}$, 
a surface gravity of log(g) = 3.5--4.5, and a temperature of $T_{eff}$ $\approx$ 1800--2600 $K$. 

\subsection{Limits on the Age and Mass of ROXs 42Bb}
Circumstantial evidence suggests that ROXs 42B's age may be between that of most $\rho$ Oph cloud members and the distributed population.
ROXs 42B appears to lie very close to or along a filamentary structure 
defining members of the ``ROXs 43A group", a sub-clump of coeval very young stars \citep{Makarov2007} located 
just outside L1989N/S.  Assuming that the brighter primary component of ROXs 42Bb 
is an M0 star and 
contributes 67\% of the total system luminosity \citep[c.f.][]{Ratzka2005}, 
adopting a distance of 135 $pc$, using the bolometric correction\footnote{Here, we convert 
from a dereddened $V$-band magnitude of $\approx$ 12.27 \citep[c.f.][]{Cieza2007,Cardelli1989}.}
and effective temperature scale appropriate for young stars from \citet{Pecaut2013}, 
we estimate an age of $\approx$ 2.5--3 $Myr$ from the \citet{Baraffe1998} isochrones.  
If instead ROXs 42Bb is an M1 (K5) star, its estimated age is 2 (6) $Myr$.
Performing this analysis for all the GKM group members located within 3 arc-minutes 
of ROXs 42B yields a median age of $\approx$ 2--4 $Myr$.

ROXs 42B could be older if it is an interloping member Upper Scorpius association 
\citep[5 or 11 $\pm$ 2 Myr][]{Preibisch2002,Pecaut2012} which is located in close proximity to $\rho$ Oph and has a nearly identical mean proper motion.  
On the other hand, ROXs 42B's position within a filamentary structure, closer proximity to both the ROXs 43A 
group and the main $\rho$ Oph cloud complexes, and extinction ($A_{V}$ $\sim$ 1.9) higher than 
all but $\approx$ 5\% of all  Upper Sco members (none of which are located close to ROXs 42B) 
\citep[c.f.][]{Carpenter2009} makes it an unlikely interloper.

We use the \citet{Spiegel2012} atmosphere/evolution models and AMES/DUSTY \citep{Allard2001} 
atmosphere models coupled with the \citet{Baraffe2003} evolution models to estimate the masses of companions 
with photometry matching that of ROXs 42Bb (Figure \ref{massratio}, right panel).  Assuming an age of ROXs 42B equal to that of the mean 
age for embedded regions of $\rho$ Oph (1--2 $Myr$) the predicted mass for ROXs 42Bb 
 ranges between $M$ $\sim$ 6 $M_{J}$ and 9 $M_{J}$.
Assuming instead that its age is more similar to that of the lower extinction, distributed population 
\citep[$\tau$ $\sim$ 3 Myr][]{Erickson2011}, the mass of the companion is 11 $M_{J}$.
If ROXs 42B is an Upper Sco member, ROXs 42Bb's estimated mass is $\approx$ 13-15 $M_{J}$.  

Considering these analyses, we conservatively estimate a mass range for ROXs 42Bb of 6--15 $M_{J}$: planet mass (9$_{-3}^{+2}$ $M_{J}$) 
if it is a bona fide member of the $\rho$ Oph complex or at/near the deuterium-burning limit 
 nominally separating planets from brown dwarfs if it is an interloping member of Upper Scorpius.

\section{Discussion}
\begin{figure}
\centering
\includegraphics[scale=0.45]{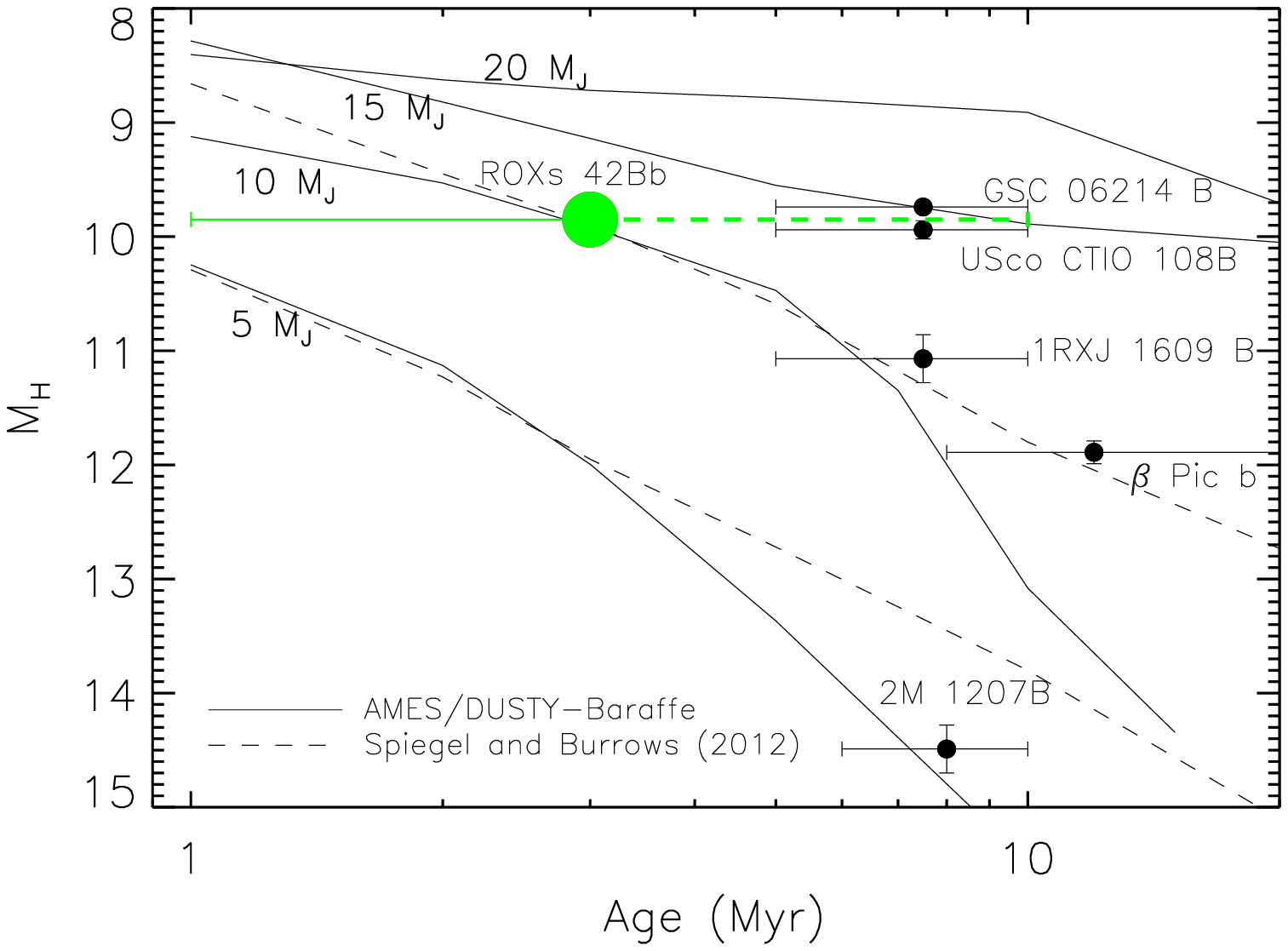}
\includegraphics[scale=0.45]{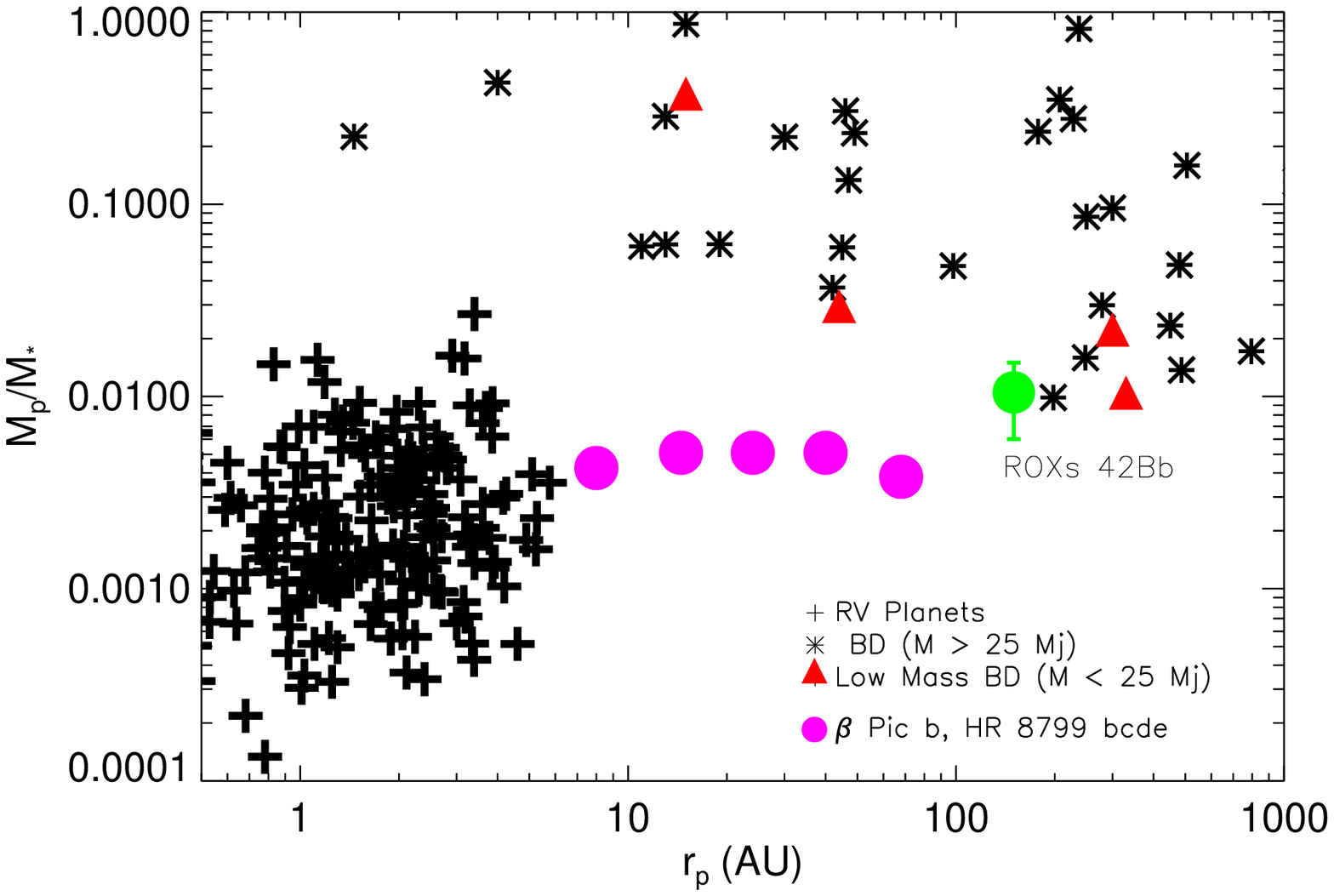}
\caption{Mass estimates for ROXs 42Bb assuming the \citet{Baraffe2003,Spiegel2012} 
luminosity evolution models (left) and its relationship to planets/brown dwarfs (right) 
assuming that ROXs 42B's primary star mass is 1 $M_{\odot}$ ($M_{prim}$ $\approx$ 0.65 + 0.35 $M_{\odot}$)).
Given the plausible age for ROXs 42Bb, its most likely mass is 6--15 $M_{J}$: 
6--11 $M_{J}$ if it is a member of $\rho$ Oph (solid horizontal error bars) and
 and 13--15 $M_{J}$ if it is an Upper Sco (dashed horizontal error bars) member.
ROXs 42Bb could be similar to low-mass brown 
dwarf companions or may fill in the gap between brown dwarfs on one side and 
radial-velocity and bona fide imaged planets on the other.
}
\label{massratio}
\end{figure}

We present evidence that ROXs 42B is likely orbited by a substellar, possibly planetary-mass companion 
at $r_{proj}$ $\approx$ 150 $AU$.  
ROXs 42Bb may be valuable in clarifying (or complicating) our understanding of the formation of 
wide-separation planets and brown dwarfs.  To illustrate, in Figure \ref{massratio} (right panel) 
we compare ROXs 42Bb's mass ratio to those for the sample of radial-velocity detected 
planets, brown dwarf companions, and directly-imaged planets around HR 8799 and $\beta$ Pic 
as in \citet{Currie2011a}.  

As argued by \citet{Currie2011a}, a gap exists between 
HR 8799 bcde and $\beta$ Pic b on one side and brown dwarf companions on the other, where 
the former (imaged planets) may be the high mass ratio, wide separation extrema of a population 
contiguous with RV-detected planets, Jupiter, and Saturn.  Similarly, \citet{Janson2012a} 
compared limits on the frequency of wide-separation companions around 
stars of different masses to the efficiency of different planet formation mechanisms, 
concluding that disk instability is unlikely to 
explain these objects.   Although core accretion is inefficient at wide separations and 
challenged in being able to form 5--10 $M_{J}$ companions, it nevertheless probably 
gave rise to planets like HR 8799 bcde and $\beta$ Pic b.

As depicted in Figure \ref{massratio}, ROXs 42Bb may somewhat blur the division between 
bona fide planets and those of the lowest mass brown dwarfs (i.e. 'planet-mass brown dwarfs').
Other means
may be required to shed light on the formation 
of wide-separation, super jovian-mass companions \citep[e.g.][]{Konopacky2013}.
The system may be especially difficult to interpret if the second companion is likewise 
bound and substellar.  Our preliminary analysis of additional data
 suggests that this ``ROXs 42C'' is brighter than ROXs 42Bb at $J$ band and fainter at 
$L^\prime$, implying significantly bluer colors indicative of a background object.  
However, new astrometry is required to 
demonstrate that it is not bound to the primaries.

We leave these issues, a more detailed investigation of the host star properties, and 
further analysis of the atmosphere of ROXs 42Bb to future studies.

\acknowledgements 
We thank Eric Mamajek, Christian Marois, Markus Janson, Rachael Friesen, Peter Plavchan, 
and the anonymous referee
 for helpful manuscript comments/suggestions and the Subaru and ESO Time Allocation Committees
 for generous allotments of observing time.  This research made use of the Keck Observatory Archive (KOA).
   TC and SD are supported by McLean Postdoctoral Fellowships.

{}

\begin{deluxetable}{lllllllll}
\tablecolumns{9}
\tabletypesize{\tiny}
\tablecaption{Observing Log and Companion Measurements}
\tablehead{{UT Date}&{Program ID}&{Telescope/Camera}&{Filter}&{$t_{int}$ ($s$)}&{$N_{images}$} & $\Delta$(mag) & 
{$r$ (\arcsec{})} & {$PA$ (deg.)}}
\startdata
\tiny
\textit{Published}$^{a}$\\
2001-07-01 & - & NTT/SHARP-I&$K$ & - & -& 6.74 $\pm$ 0.54& 1.137 $\pm$ 0.014 & 268.0 $\pm$ 0.3\\
\textit{Archival}$^{b}$\\
2005-04-16 & U55N2&Keck/NIRC2 & $K$ & 7.5 & 6 & 6.30 $\pm$ 0.15 & 1.157 $\pm$ 0.010 & 268.88 $\pm$ 0.60\\
2008-07-19 & S08A-074&Subaru/CIAO & $K_{s}$ & 9.9 & 5 & 6.10 $\pm$ 0.20 & 1.160 $\pm$ 0.010 & 269.67 $\pm$ 1.00\\
2011-06-22 & H242N2&Keck/NIRC2 & $H$ & 10 & 8 & 6.86 $\pm$ 0.05 & 1.169 $\pm$ 0.005 & 269.65 $\pm$ 0.25\\
2012-08-11 & 089.C-0411(A)&VLT/SINFONI & $K$ IFS & 20 & 3 & 6.2 $\pm$ 0.2 & 1.15 $\pm$ 0.03 & 270 $\pm$ 2
 \enddata
\tablecomments{a) - Presented in \citet{Ratzka2005}. b) Data we present for the first time.}
\label{obslog}
\end{deluxetable}

\end{document}